\documentclass[12pt]{article} 

\usepackage{ol2}

\usepackage[draft]{hyperref}
\usepackage{amsmath}

\usepackage{color} 

\begin{document}


\title{25-nm diamond crystals hosting single NV color centers sorted by photon-correlation near-field microscopy}

\author{Yannick Sonnefraud$^{1}$, Aur\'{e}lien Cuche$^{1}$, Orestis Faklaris$^{2}$,
Jean-Paul Boudou$^{3}$, Thierry Sauvage$^{4}$, Jean-Fran\c{c}ois Roch$^2$, Fran\c{c}ois Treussart$^{2}$,
Serge Huant$^{1,*}$}

\address{$^{1}$ Institut N\'{e}el, CNRS and Universit\'{e} Joseph Fourier, BP166, 38042 Grenoble Cedex, France \\
$^{2}$ Laboratoire de Photonique Quantique et Mol\'{e}culaire, UMR 8537 CNRS/\'Ecole Normale Sup\'erieure de Cachan, France\\
$^{3}$ BioEmco, UMR 7618 CNRS et Universit\'{e} Pierre et Marie Curie, Paris, France\\
$^{4}$ Centre d'\'{E}tude et de Recherche par Irradiation, CNRS, Orl\'{e}ans, France\\
$^{*}$ Corresponding author : serge.huant@grenoble.cnrs.fr }

\begin{abstract}
Diamond nanocrystals containing highly photoluminescent color centers are attractive non-classical and near-field light sources. For near-field applications the size of the nanocrystal is crucial since it defines the optical resolution. NV (Nitrogen-Vacancy) color centers are efficiently created by proton irradiation and annealing of a nanodiamond powder. Using near-field microscopy and photon statistics measurements, we show that nanodiamond with size down to 25~nm can hold a single NV color center with bright and stable photoluminescence.
\end{abstract}


Probing electromagnetic field at the nanoscale level is a challenging field of research~\cite{NovotnyBook2006}. Recently, single fluorescent molecules at room temperature have been used to probe the near field of a resonant antenna positionned at the end of a metal-coated glass fiber near-field probe~\cite{Taminiau2007}.
Fluorescent-active tips consisting of a single fluorescent nano-object attached to a tip offer many prospects in Near-Field Scanning Optical Microscopy (NSOM) since the spatial resolution is expected to scale down to a dimension approaching the size of the emitter, i.e., around 10~nm for a single CdSe nanocrystal~\cite{Chevalier2005}. Moreover, such active tip is a unique technique to study the interaction of a nano-object with its environment~\cite{Sandoghdar2005}.
This should open further perspectives in nano-optics and quantum-optics experiments.

Several kinds of fluorescent-active objects have been considered such as a single terrylene molecule in a p-terphenyl micro-crystal~\cite{Michaelis2000}, rare-earth-doped glass particles~\cite{Aigouy2003}, CdSe nanocrystals~\cite{Shubeita2003,Chevalier2005,Sonnefraud2006}, color centers in a thin LiF layer~\cite{Sekatskii2007} and NV color centers in a diamond nanocrystal~\cite{Kuhn2001}. So far, none of these has revealed satisfactory for a practical use in NSOM. Although near-field imaging using single molecules at cryogenic temperatures is very successful~\cite{Michaelis2000}, it can hardly be extended at room temperature due to photobleaching~\cite{Jacques2007}.
Experiments using CdSe nanocrystals offered insufficient control over the very object attached to the tip~\cite{Chevalier2005} and optical imaging with them turned out to be restricted by blinking~\cite{Sonnefraud2006}. In addition, the achieved spatial resolution is similar to conventional NSOM due to either the large size of the material hosting the emitter or/and the number of active centers that is required for optical detection.

Ideally a fluorescent nano-object should combine several properties for application as a practical active tip: it should emit in ambient conditions, have a high quantum efficiency, exhibit a radiative lifetime short enough to allow for a large counting rate at the single emitter level; it should neither blink nor bleach over a long measurement duration; and it should be hosted in a material of nanometric dimension as small as possible. The nano-object should eventually host a single emitter with known lifetime and dipole orientation.

NV$^-$ color centers in diamond~\cite{Gruber1997} combine many of these demanding characteristics. Their photoluminescence centered at 670~nm at room temperature has a near-unity quantum efficiency and corresponds to a radiative lifetime of 11~ns. They are extremely stable emitters: no blinking, nor bleaching was reported so far for a single emitter.
Photon antibunching~\cite{Beveratos2001} and single-photon emission~\cite{Beveratos2002} have been demonstrated for a NV color center in a single nanodiamond. Such photoluminescent nanoparticles have already been successfully used as nanoscopic light source for NSOM~\cite{Kuhn2001} but the achieved resolution was about 300~nm, a value presumably limited by the size of the crystal attached to the tip. Significant improvement in resolution requires much smaller photoluminescent nanodiamonds compared to this pioneering experiment.

In this letter, we demonstrate that a single NV color center can be detected in individual nanodiamonds with size smaller than 25~nm in-diameter. This is achieved by combining a NSOM, which offers a dual topographical and optical operating mode, with a photon-correlation measurement. 
This combination gives access in a single setup to the size of the particle, its emission ability, the number of active emitters and their photoluminescence spectrum. We confirm that single NV centers in such small nanodiamonds neither blink nor bleach under cw excitation over an observation duration longer than one hour.

The experiment is based on a home-made NSOM~\cite{Sonnefraud2006}. A Hanbury Brown and Twiss (HBT) intensity correlator is added to the setup: the light is collected from the sample in transmission through a high-numerical aperture microscope objective. It is then guided to the HBT correlator by an optical fibre, recoupled to free space, separated by a 50/50 beamsplitter and finally focused on two avalanche photodiodes (APDs, SPCM-AQR~13, Perkin-Elmer, Canada) connected to a Time Correlated Single-Photon Counting module (PicoHarp~300, PicoQuant, Germany). A glass filter and a diaphragm are placed in front of each APD to eliminate optical cross-talk between the detectors~\cite{Kurtsiefer2001}.
The NSOM tip is produced from a single-mode optical fibre with a pure silica core chemically etched in HF. The tip is uncoated allowing us to inject a large optical power without subsequent damaging of the tip. In addition, the very small radius of curvature at the apex (down to 30~nm) achieves higher topographical resolution compared to a blunt metal-coated tip, though at the expense of reduced optical resolution. The near-field optical resolution is about 400~nm, a value slightly better than the 470~nm one achieved in the confocal mode.

Nanodiamonds are produced as follows. A commercial high-pressure high-temperature synthetic diamond powder (SYP 0-0.05, Van Moppes, Switzerland) of type Ib ($\approx 100$~ppm nitrogen content), with particle size below 50~nm, is proton-irradiated at 2.5~MeV, 5.10$^{15}$~H$^+$.cm$^{-2}$ in a Van de Graaff accelerator. The irradiation efficiently generates vacancies in the diamond lattice.  NV centers are created by annealing (800$^\circ$C, 2~h) which activates migration of vacancies resulting in their trapping by the nitrogen impurities. Remaining nitrogen atoms act as donors stabilizing the negative charge state of the NV$^-$ center~\cite{Davies1977}. The now optically-active nanodiamonds are acid treated (H$_2$SO$_4$:HNO$_3$, 9:1 vol:vol, at 75$^\circ$C for 3 days, following a procedure described in~\cite{Fu2007}), and washed with water, leading to a stable aqueous suspension thanks to repulsive electrostatic interactions between charged surface chemical groups. Intense ultra-sonification is used to disperse the particles at their primary size. Finally, polyvinylalcool (PVA) is added to the solution to reach a proportion of 0.3\% in weight.

A 10~$\mu\ell$ droplet is then spin cast on a fused silica microscope coverslip. The tip is approached to the sample, shear-force feedback keeping tip-sample distance constant. Images are reconstructed by raster scanning the tip over the sample. The 488~nm line of an Ar-Kr cw laser is coupled directly to the tip and luminescence of the NV color centers is excited by the light funnelled through its apex. The photoluminescence is collected by the microscope objective and spectrally filtered from the excitation light by a dichroic mirror and a bandpass filter (605-755~nm). This filter, adapted to the NV$^-$ center photoluminescence, partially cuts the neutral NV$^0$ centers emission~\cite{Dumeige2004}. Light coming from the excitation volume defined by the tip aperture is coupled to an optical fibre (50~$\mu$m core diameter) and is guided after spectral filtering either to a single APD for imaging, or to a spectrograph, or to the HBT intensity correlator.

\begin{figure*}[htb] 
\begin{center}
\includegraphics[width=8.3cm]{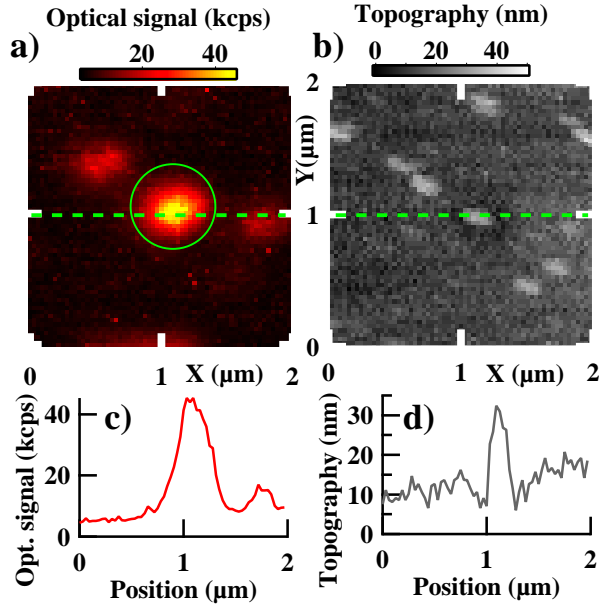}
\end{center}
 \caption{\label{figure1}NSOM detection of a single NV color center in a nanodiamond. \textbf{(a)} Luminescence image recorded with an excitation optical power of 100~$\mu$W at the tip apex. Integration time: 140~ms per pixel, scan speed: 0.5~$\mu$m.s$^{-1}$. \textbf{(b)} Corresponding topography. The image has been numerically flattened. \textbf{(c)}~(resp. \textbf{(d)}) Line cut along the green dashed line of images (a) (resp. (b)).}
\end{figure*} 

\begin{figure*}[htb] 
\begin{center}
\includegraphics[width=8.3cm]{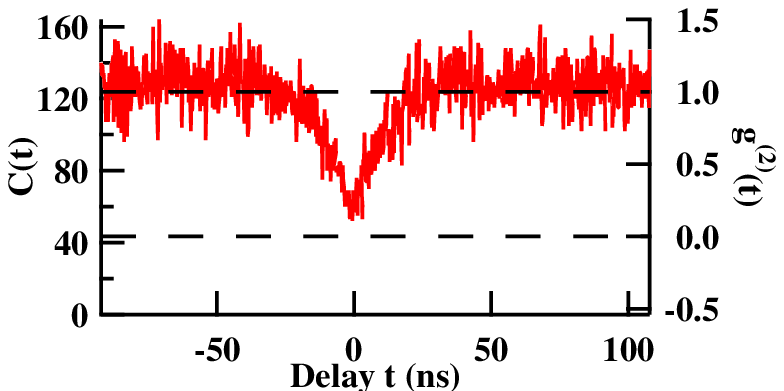}
\end{center}
 \caption{\label{figure2}Photon antibunching giving evidence for emission of a single NV center. Right scale: normalized correlation function $g^{(2)}(t)$  for the NV color center circled in Fig.\ref{figure1}.a. Raw coincidence numbers are indicated on the left axis. The $g^{(2)}(t)$ function is evaluated by accounting for the background contribution in the normalization of $C(t)$.}
\end{figure*} 

To first localize the nanoparticles, NSOM imaging is performed on a selected region of the sample, as displayed on figures~\ref{figure1}.a and figure~\ref{figure1}.b. The topographic signal in Fig.\ref{figure1}.b enables a precise localisation of the particles and the luminescence image in Fig.\ref{figure1}.a reveals those which are optically active, i.e., which host NV centers.

A line cut along the topographic signal reveals that the particle at the center of the images is as small as 25~nm in height, with larger lateral dimensions due to convolution between the tip apex and the particle. The topography image shows a small ring-shaped gap around the particle, which is tentatively attributed to a depletion of the PVA layer.

Although we scanned a limited number of images, we found that 24 particles out of 85, i.e. 28~\%, were photoluminescent and did not show any blinking. In addition, among the optically active nanodiamonds, 6 (respectively 8) had diameters around 20~nm (25~nm). Some of them were even smaller: as an example, the photoluminescent middle right nanodiamond on Fig.\ref{figure1}.a has a diameter around 15~nm.

Once an emitting particle is detected, we place the tip in front of it and plug the detection fibre to the HBT setup. The correlation histogram of Fig.\ref{figure2}, $C(t)$, corresponding to the encircled object of figure~\ref{figure1}.a, exhibits a dip at zero delay, which falls down to a value lower than half its long-delay limit. This antibunching effect already proves that the collected light is associated to emission by a single color center. 

To infer the second-order correlation function $g^{(2)}(t)$, $C(t)$ is first normalized as $C_N(t)=C(t)/(wN_1N_2T)$, with $w=256~$ps the time bin of the correlation histogram, $N_1=23$~kcps and $N_2=2$~kcps are the count rate at each detector in counts per seconds (cps), and $T=1000~$s the total integration time~\cite{Brouri2000}. $g^{(2)}(t)$ is then obtained by correcting $C_N(t)$ from the background light using the relation $g^{(2)}(t)=[C_N(t)-(1-\rho^2)]/\rho^2$, where $\rho=S/(S+B)$ is associated to the signal($S$)-to-background($B$) ratio. 
This parameter is evaluated on the optical NSOM image using line cuts shown in Fig.\ref{figure1}.d. For the present object, $B$~=6~kcps, $S$=38.5~kcps and  $\rho$~=~0.80.
The correlation function shows an anti-bunching gap at zero delay with $g^{(2)}(0)$~=~0.1. This unambiguously confirms that we indeed address a single NV color center~\cite{Beveratos2001, non_zero_g2}.

We have analyzed a few other emitting nanodiamonds of similar size and found that most of them host one and sometimes two color centers, like the upper left 30~nm nanodiamond on Fig.~\ref{figure1}.a.  Rabeau \textit{et al.}~\cite{Rabeau2007} combined atomic force and confocal microscopy to correlate the fluorescence from CVD-grown individual nanodiamonds to their size. Under the CVD growth conditions, they found an optimal size of 60-70~nm for a single NV center occupancy and they observed that particles smaller than about 40~nm are not photoluminescent. Using irradiation of nanodiamond powder, we observe emission from NV color centers in particles as small as 15~nm.

In summary, we have combined NSOM with photon-correlation measurement to detect single NV$^-$ color centers in nanodiamonds smaller than 25~nm in-diameter. This powerful combination permits a complete topographical and optical sorting of the nanodiamonds. The associated photoluminescence is remarkably stable without any blinking. Since we are working to further reduce the size of these particles and to increase their color centers content, they show great promise for a wide range of future applications in biology~\cite{Fu2007}, such as marking and long-term tracking, and in nano-optics to realize NSOM with optically active tips.\smallskip \\  

We are grateful to Fedor Jelezko and Jean-Philippe Poizat for very fruitful discussions. This work is partly supported by the European Commission through the \emph{Nano4Drugs} (contract LSHB-2005-CT-019102), and the \emph{EQUIND} (contract IST-034368) projects. AC aknowledges funding by the R\'egion Rh\^one-Alpes through the ''Cluster Micro-Nano''.




\begin{thebibliography}{99}
\bibitem{non_zero_g2} The non-zero $g^{(2)}(0)$ value is attributed to background emission from the nanocrystal itself which cannot be separately measured from the color center luminescence.
\end{thebibliography}
\end{document}